\documentclass[aps,prl,reprint,preprintnumbers,superscriptaddress,amsmath,amssymb,bibnotes,longbibliography]{revtex4-2}
\usepackage{graphicx}
\usepackage{dcolumn}
\usepackage{bm}
\usepackage[colorlinks,linkcolor=blue,anchorcolor=blue,citecolor=blue,breaklinks,CJKbookmarks=True,urlcolor=blue,filecolor=blue,menucolor=blue,runcolor=blue]{hyperref}

\begin{document}

\title{Structural phase transitions and superconductivity in the Heusler intermetallics $X$Pd$_2$Sn ($X$ = Ti, Zr, Hf)}

\author{Hang Su}
\affiliation  {Center for Correlated Matter and Department of Physics, Zhejiang University, Hangzhou 310058, China}
\affiliation  {Zhejiang Province Key Laboratory of Quantum Technology and Device, Department of Physics, Zhejiang University, Hangzhou 310058, China}
\author{Feng Du}
\affiliation  {Center for Correlated Matter and Department of Physics, Zhejiang University, Hangzhou 310058, China}
\affiliation  {Zhejiang Province Key Laboratory of Quantum Technology and Device, Department of Physics, Zhejiang University, Hangzhou 310058, China}
\author{Rui Li}
\affiliation  {Center for Correlated Matter and Department of Physics, Zhejiang University, Hangzhou 310058, China}
\affiliation  {Zhejiang Province Key Laboratory of Quantum Technology and Device, Department of Physics, Zhejiang University, Hangzhou 310058, China}
\author{Shuaishuai Luo}
\affiliation  {Center for Correlated Matter and Department of Physics, Zhejiang University, Hangzhou 310058, China}
\affiliation  {Zhejiang Province Key Laboratory of Quantum Technology and Device, Department of Physics, Zhejiang University, Hangzhou 310058, China}
\author{Yuxuan Chen}
\affiliation  {Hangzhou Xuejun High School, Hangzhou 310012, China}
\author{Jiyong Liu}
\affiliation  {Department of Chemistry, Zhejiang University, Hangzhou 310027, China}
\author{Ye Chen}
\affiliation  {Center for Correlated Matter and Department of Physics, Zhejiang University, Hangzhou 310058, China}
\affiliation  {Zhejiang Province Key Laboratory of Quantum Technology and Device, Department of Physics, Zhejiang University, Hangzhou 310058, China}
\author{Chao Cao}
\affiliation  {Center for Correlated Matter and Department of Physics, Zhejiang University, Hangzhou 310058, China}
\author{Michael Smidman}
\email[Corresponding author: ]{msmidman@zju.edu.cn}
\affiliation  {Center for Correlated Matter and Department of Physics, Zhejiang University, Hangzhou 310058, China}
\affiliation  {Zhejiang Province Key Laboratory of Quantum Technology and Device, Department of Physics, Zhejiang University, Hangzhou 310058, China}
\author{Huiqiu Yuan}
\email[Corresponding author: ]{hqyuan@zju.edu.cn}
\affiliation  {Center for Correlated Matter and Department of Physics, Zhejiang University, Hangzhou 310058, China}
\affiliation  {Zhejiang Province Key Laboratory of Quantum Technology and Device, Department of Physics, Zhejiang University, Hangzhou 310058, China}
\affiliation  {State Key Laboratory of Silicon Materials, Zhejiang University, Hangzhou 310058, China}
\affiliation  {Collaborative Innovation Center of Advanced Microstructures, Nanjing University, Nanjing, 210093, China}

\date{\today}


\begin{abstract}
We report the discovery of structural phase transitions and superconductivity in the full Heusler compounds $X$Pd$_2$Sn ($X$ = Ti, Zr, Hf), by means of electrical transport, magnetic susceptibility, specific heat and x-ray diffraction measurements. TiPd$_2$Sn, ZrPd$_2$Sn and HfPd$_2$Sn undergo structural phase transitions from the room-temperature cubic MnCu$_2$Al-type structure (space group $Fm\bar{3}m$) to a low-temperature tetragonal structure at around 160~K, 110~K and 90~K, respectively, which are likely related charge density wave (CDW) instabilities. Low temperature single crystal x-ray diffraction measurements of ZrPd$_2$Sn demonstrate the emergence of a superstructure with multiple commensurate modulations below $T_s$. ZrPd$_2$Sn and HfPd$_2$Sn have bulk superconductivity (SC) with transition temperatures $T_c$ $\sim$ 1.2~K and 1.3~K, respectively. Density functional theory (DFT) calculations reveal evidence for structural and electronic instabilities which can give rise to CDW formation, suggesting that these $X$Pd$_2$Sn systems are good candidates for examining the interplay between CDW and SC.

\end{abstract}

\maketitle


\section{\uppercase\expandafter{\romannumeral1}. INTRODUCTION}

Since the discovery of the prototype compound MnCu$_2$Al by Heusler in 1903, a large number of Heusler intermetallic materials have been reported with the formula $AT_2M$ \cite{heusler1903, dshemuchadse2015more, pottgen2019intermetallics}, where $A$ is an early transition metal or small rare earth, $T$ is a late transition metal and $M$ is an electronegative p-block element. Most Heusler compounds have a cubic structure and a wide variety of physical properties and ground states have been reported, including heavy fermion behavior~\cite{TAKAYANAGI1988281, GOFRYK2005625}, shape memory effects~\cite{LiuAPL2003}, topologically nontrivial states~\cite{GuoPhysRevB2017} and half-metallic ferromagnetism~\cite{ChenAPL2006, kourov2015specific, ShanPhysRevLett2009}. Due to the large number of members in this family, many of the properties can be tuned by chemical substitution \cite{Wollmann2017, Alijani2011, GRAF20111}. In addition, a tetragonal crystal structure has been observed in a few Heusler compounds [e.g. $T$Rh$_2$Sn ($T$ = V, Cr, Fe, Co)], which corresponds to a distortion of the typical cubic structure \cite{SUITS1976}. Superconductivity (SC) has also been reported in Heusler materials, mostly in compounds containing Pd, Ni, or Au at the $T$ site and the highest observed transition temperature $T_c$ is currently 4.9~K (YPd$_2$Sn) \cite{WERNICK198390}. In addition, the coexistence of superconductivity (SC) and long range magnetic order has been observed in YbPd$_2$Sn and ErPd$_2$Sn~\cite{shelton1986coexistence}. The valence electron count (VEC) has been found to be closely related to the superconducting properties, where the maximum values of $T_c$ are at around VEC = 27 \cite{KlimczukPhysRevB}.

Charge density wave (CDW) order corresponds to an instability in the electronic states near the Fermi level, where there is a periodic modulation of the electronic charge below the transition temperature due to the opening of a gap, which is usually concomitant with a structural phase transition \cite{McMillanPhysRevB1976, GABOVICHphysicsreport2002}. The coexistence of CDW and SC has been reported in some systems and the interplay between these orders has been of great interest \cite{gruner1988dynamics, ChenROPP2016, KusmartsevaPhysRevLett2009, KlintbergPhysRevLett2012, OrtizPhysRevMaterials2019, OrtizPhysRevLett2020, DuPhysRevB135, DuFeng135CPB, ShenBPhysRevB2020, DuFPhysRevB2020}. The Heusler compound LuPt$_2$In was recently reported to exhibit a second-order CDW transition at $T_{\mathrm{CDW}}$ = 490~K, as well as superconductivity below $T_c$ = 0.45~K. Substituting Pd for Pt suppresses $T_{\mathrm{CDW}}$ to zero giving rise to a possible quantum critical point (QCP). $T_c$ was also strongly enhanced as the CDW is suppressed, giving rise to a new scenario for the interplay between CDW and SC \cite{gruner2017charge}. On the other hand besides LuPt$_2$In, structural distortions are also reported in YbPt$_2$In \cite{GrunerIOP2014} and REAu$_2$In (RE = La-Nd) \cite{BESNUS1986ReAUIn}, but the coexistence of CDW/structural phase transitions and SC is not commonly observed in Heusler materials. It is therefore of particular interest to identify additional Heusler compounds exhibiting both CDW order and superconductivity, in order to examine the relationship between these competing orders, determine the origin of the CDW instabilities and to search for potential CDW QCP's.

In this article, we report the discovery of structural phase transitions and superconductivity in the Heusler compounds $X$Pd$_2$Sn ($X$ = Ti, Zr, Hf). Electrical resistivity, specific heat, magnetic susceptibility and x-ray diffraction measurements indicate that TiPd$_2$Sn, ZrPd$_2$Sn and HfPd$_2$Sn undergo CDW-like structural phase transitions from the cubic MnCu$_2$Al type phase to a tetragonal structure upon cooling across $T_s$ $\sim$ 160~K, 110~K and 90~K, respectively. The structural instabilities are also revealed by DFT calculations of the imaginary phonon modes. At lower temperatures, ZrPd$_2$Sn and HfPd$_2$Sn undergo superconducting transitions at $T_c$ $\sim$ 1.2~K and 1.3~K, respectively. These findings suggest that $X$Pd$_2$Sn are good candidates for investigating the interplay of CDW order and SC.

\section{\uppercase\expandafter{\romannumeral2}. EXPERIMENTAL METHODS}

Polycrystalline samples of $X$Pd$_2$Sn ($X$ = Ti, Zr, Hf) were prepared by arc-melting the pure metals under an argon atmosphere. The samples were sealed in evacuated quartz tubes and annealed at 850 $^{\circ}$C for two weeks to ensure homogeneity. The crystal structures were characterized by powder x-ray diffraction (XRD) on a PANalytical X'Pert MRD diffractometer with Cu K$\alpha$ radiation, including low temperature measurements down to 30~K. The electrical transport and specific heat measurements were performed on a Physical Property Measurement System (PPMS, Quantum Design) with a $^3$He insert. The magnetization was measured using a SQUID magnetometer (MPMS-5T, Quantum Design). Single crystal XRD measurements were performed down to 90~K using a Bruker D8 Venture diffractometer. Tiny single crystals were selected from surface of the polycrystalline samples. The reciprocal space reconstruction of the single-crystal XRD data was performed using the APEX4 software. The electronic structure and density of states (DOS) of $X$Pd$_2$Sn ($X$ = Ti, Zr, Hf) were calculated using density functional theory (DFT) as implemented in VASP \cite{KressePhysRevB1993, KressePhysRevB1999}. The Perdew-Burke-Ernzerhoff (PBE) flavor of the generalized gradient approximation was employed, where the energy cutoff was chosen to be 400~eV and the Brillouin zone was sampled with 12$\times$12$\times$12 $\Gamma$-centered K-mesh. Spin-orbit coupling was taken into account using a second variation method.
The first-principles results were fitted to a tight-binding model Hamiltonian using maximally projected Wannier functions (MLWF) to calculate the static bare electron susceptibility \cite{SouzaPhysRevB2001, ArashMOSTOF2008}.
The phonon dispersion was calculated using density functional perturbation theory (DFPT) as implemented in the Quantum ESPRESSO package \cite{Giannozzi2009JPCM}. The DFPT calculations were performed using Troullier-Martins norm-conserving pseudopotentials, with an energy cut-off of 144 Ry and 6$\times$6$\times$6 q-mesh.

\section{\uppercase\expandafter{\romannumeral3}. RESULTS}

\subsection{\uppercase\expandafter{a}. Structural phase transitions }

  \begin{figure}
  	\includegraphics[angle=0,width=0.49\textwidth]{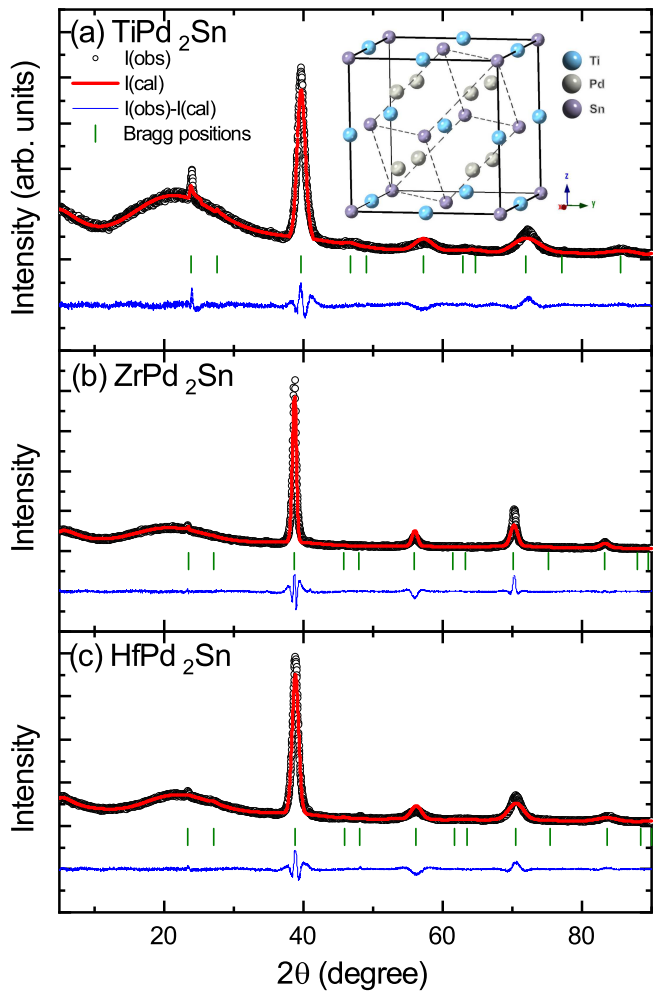}
  	\vspace{-12pt} \caption{\label{Figure1}(Color online) Room-temperature powder x-ray diffraction patterns of polycrystalline (a) TiPd$_2$Sn (b) ZrPd$_2$Sn and (c) HfPd$_2$Sn. The open symbols represent the observed data, the red solid lines denote the calculated patterns from the Rietveld refinements and the blue solid lines are the differences between the measured and calculated results. The green vertical lines denote the expected Bragg peak positions. Inset of (a): Crystal structure of TiPd$_2$Sn at room temperature. The dashed lines denote the primitive unit cell.}
  \end{figure}

  \begin{figure*}
  	\includegraphics[angle=0,width=0.98\textwidth]{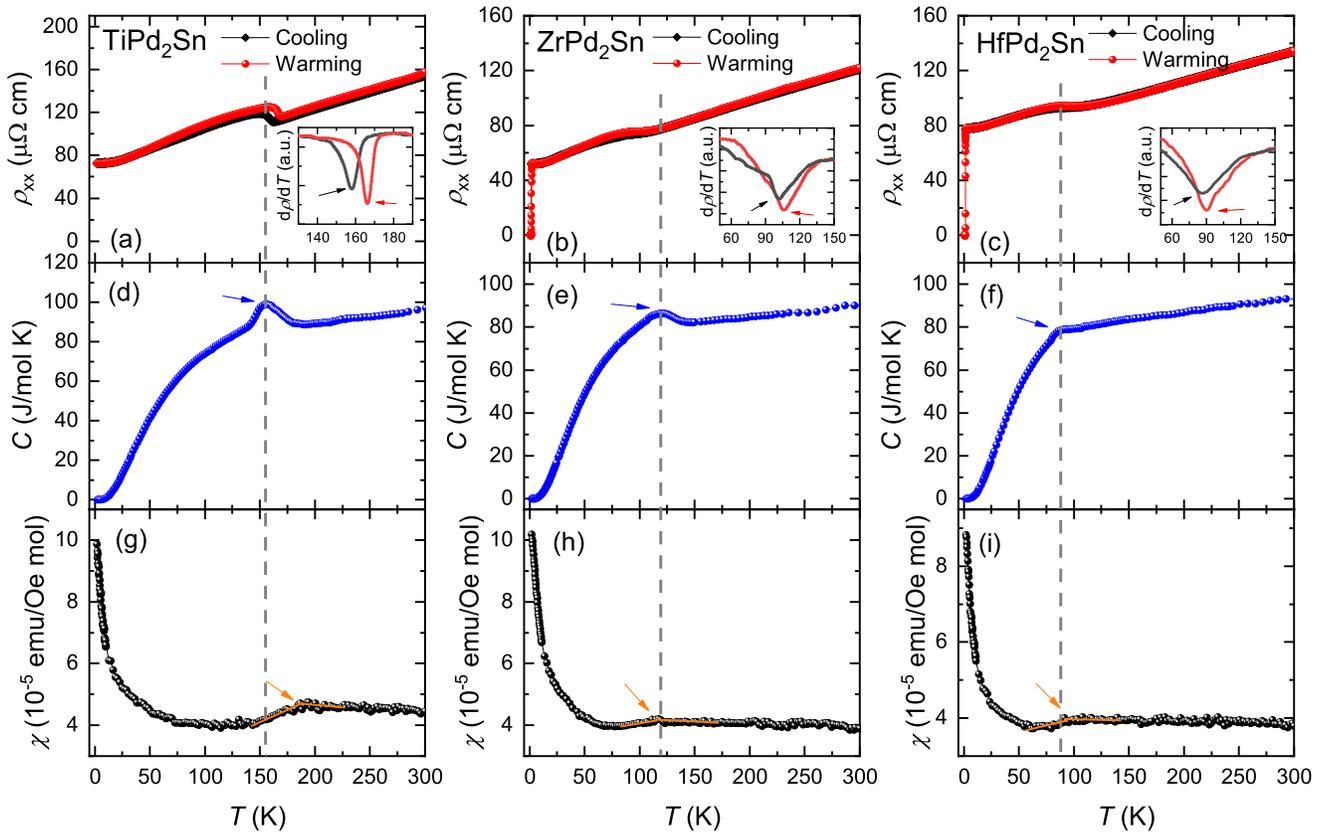}
  	\vspace{-12pt} \caption{\label{Figure2}(Color online)  Temperature dependence of (a), (b) and (c): electrical resistivity $\rho_{xx}(T)$ measured upon warming and cooling, (d), (e) and (f): specific heat $C(T)$ and (g), (h) and (i): magnetic susceptibility $\chi(T)$ for TiPd$_2$Sn, ZrPd$_2$Sn and HfPd$_2$Sn, respectively. The $C(T)$ and $\chi(T)$ data are measured upon warming. The insets show the derivative of $\rho_{xx}(T)$ near $T_s$. The arrows in all the panels denote the structural transitions (see text), corresponding to the values in Table. \ref{table:table1}. The vertical dashed lines correspond to the structural transition temperatures from the specific heat, which are shown as guides to the eye. }
  \end{figure*}

  \begin{figure*}
  	\includegraphics[angle=0,width=0.98\textwidth]{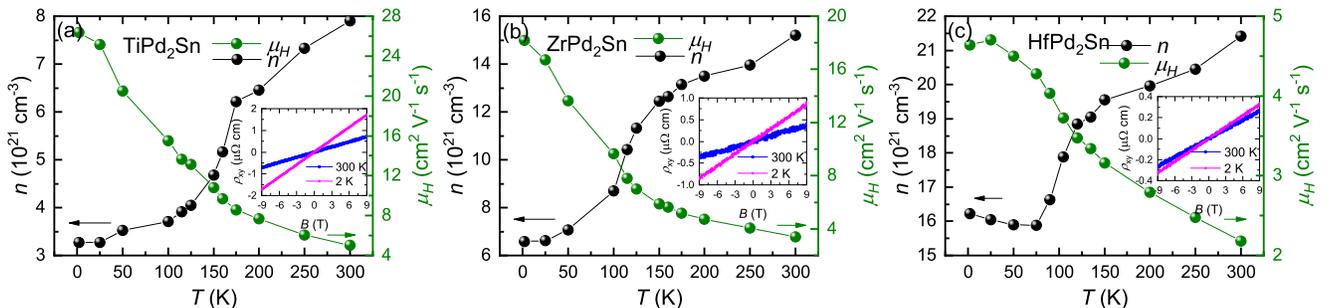}
  	\vspace{-12pt} \caption{\label{Figure3}(Color online) The temperature dependence of the carrier density and mobility of the Hall resistivity of (a) TiPd$_2$Sn, (b) ZrPd$_2$Sn and (c) HfPd$_2$Sn. The field-dependence of the Hall resistivity at two temperatures are shown in the insets. }
  \end{figure*}

  \begin{figure*}
  	\includegraphics[angle=0,width=0.98\textwidth]{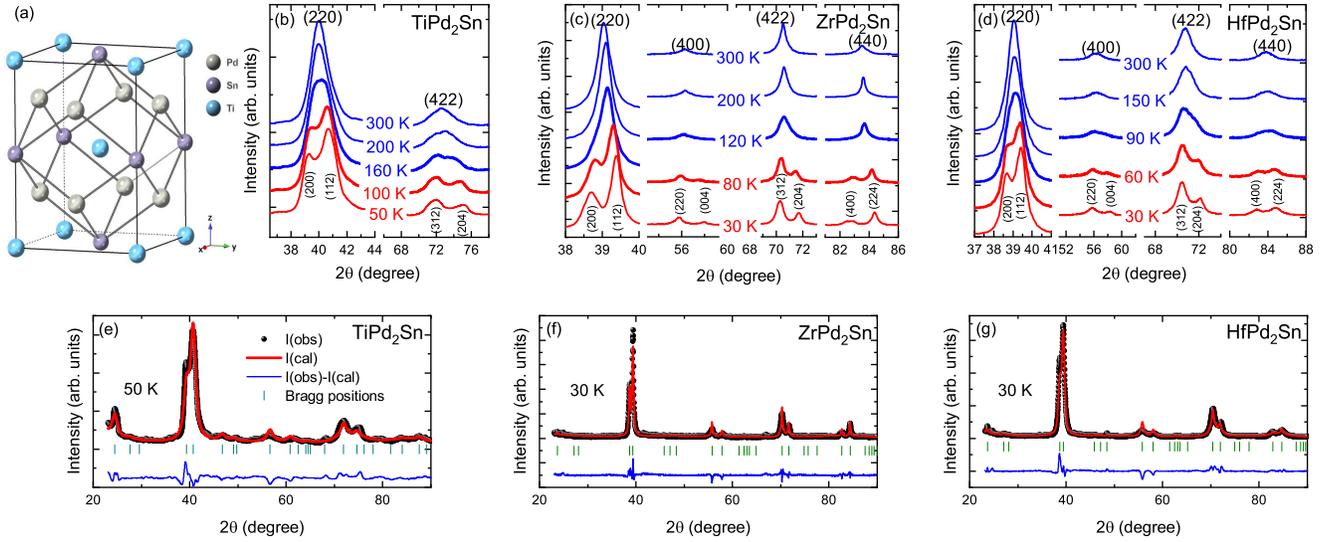}
  	\vspace{-12pt} \caption{\label{Figure4}(Color online) (a) Low temperature tetragonal crystal structure of TiPd$_2$Sn. (b), (c) and (d): powder x-ray diffraction patterns of $X$Pd$_2$Sn ($X$ = Ti, Zr, Hf) measured at various temperatures, focusing on a few major diffraction peaks. The corresponding powder x-ray diffraction patterns together with the calculated results from Rietveld refinements at the lowest measured temperatures ($T<T_s$) are shown in (e), (f) and (g).
  }
  \end{figure*}

  \begin{figure*}
  	\includegraphics[angle=0,width=0.98\textwidth]{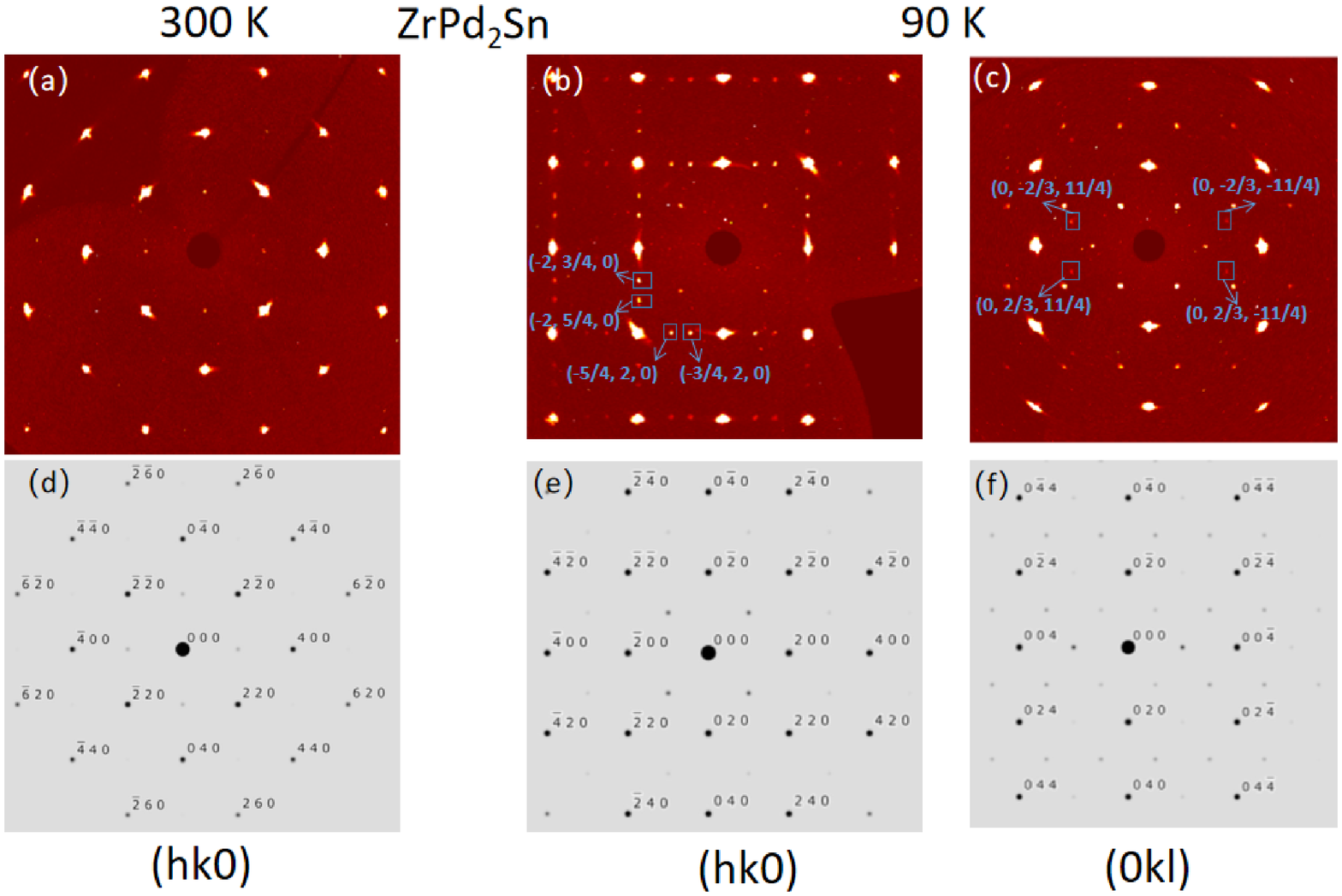}
  	\vspace{-12pt} \caption{\label{Figure45}(Color online) Single-crystal XRD results for ZrPd$_2$Sn. Reciprocal space images are shown for (a) the ($hk0$) plane at 300~K ($T>T_s$), (b) ($hk0$) and (c) ($0kl$) at 90~K ($T<T_s$). (d) Simulated diffraction pattern in the ($hk0$) plane for the cubic crystal structure (space group $Fm\bar{3}m$.) (e) and (f) Simulated diffraction pattern in the ($hk0$) and ($0kl$) planes for the low temperature tetragonal crystal structure (space group $I4/mmm$.) Representative additional low temperature satellite peaks in (b) and (c) are also labelled.
  }
  \end{figure*}

Powder XRD patterns of $X$Pd$_2$Sn ($X$ = Ti, Zr and Hf) measured at 300~K are shown in Fig. \ref{Figure1}. The patterns exhibit broadened Bragg peaks, which may be due to the small size of the crystalline grains or a reduced crystallinity. The full width at half maximum (FWHM) values of the main (220) peak are 1.25$^\circ$, 0.78$^\circ$ and 0.91$^\circ$ for TiPd$_2$Sn, ZrPd$_2$Sn and HfPd$_2$Sn, respectively. All the diffraction peaks can be well indexed to the cubic MnCu$_2$Al-type structure with space group $Fm\bar{3}m$. Rietveld refinements yield lattice parameters of $a$ = 6.3351(3)~\AA, 6.5608(2)~\AA~and 6.5272(3)~\AA~for TiPd$_2$Sn, ZrPd$_2$Sn and HfPd$_2$Sn, respectively, which are close to the previously reported values \cite{YIN201515}. The crystal structure of TiPd$_2$Sn is shown in the inset of Fig. \ref{Figure1}(a). The Ti-Pd atoms form a NaCl-type sublattice while Pd atoms fill the tetrahedral voids.

Figure \ref{Figure2} shows the temperature dependence of the electrical resistivity $\rho_{xx}(T)$, specific heat $C(T)$ and magnetic susceptibility $\chi(T)$ of the three materials, measured from 300~K to 0.5~K. Upon lowering the temperature from 300~K, the decrease of $\rho_{xx}(T)$ indicates a metallic nature. The $\rho_{xx}(T)$ of TiPd$_2$Sn exhibits a jump at around 160~K with clear hysteresis between the measurements performed upon cooling down and warming up, indicating a first-order transition. In the case of ZrPd$_2$Sn and HfPd$_2$Sn, pronounced humps and moderate hysteresis are also observed, at around 110~K and 90~K, respectively. The increase of the resistivity below $T_s$ could be due to the opening of a partial gap on the Fermi surface associated with the formation of CDW order.
In addition, $C(T)$ shown in Figs. \ref{Figure2}(d), (e) and (f) have $\delta$-like anomalies at $T_s$, with relatively broad transition widths. The temperature dependence of the magnetic susceptibility $\chi(T)$ was measured in an applied field of 4~T (Figs. \ref{Figure2}(g), (h) and (i)). Below $T_s$, there is a slight but observable drop in $\chi(T)$, in line with a decrease of the density of states at the Fermi level.
We define the resistivity transition temperature $T_s$ as where the derivative of the $\rho_{xx}(T)$ reaches a minimum (insets of Figs. \ref{Figure2}(a)-(c)). The corresponding values of $T_s$ from the specific data are determined from the peak maximum or turning point, while for the magnetic susceptibility it corresponds to the onset of the decrease.
It can be clearly seen that in all three compounds, the $T_s$ determined from $\rho_{xx}(T)$, $C(T)$, and $\chi(T)$ are close, as shown in Table. \ref{table:table1}, and do not appear to depend on the applied field.

The Hall resistivity $\rho_{xy}$ was measured at various temperatures from 300~K to 2~K as a function of applied magnetic field up to 9~T, as shown for two temperatures in the insets of Fig. \ref{Figure3}. The almost linear dependence of $\rho_{xy}(B)$ within the measured field range and the positive slope suggest that hole-type carriers are dominant for all three compounds. The carrier concentration $n$ and Hall mobility $\mu_H$ can be deduced from a one-band model and their temperature dependences are shown in Fig. \ref{Figure3}.
Here $n$ was determined from the Hall coefficient $R_H$ = 1/$ne$, where $R_H$ is derived from the slope of $\rho_{xy}$, while $\mu_H$ was determined from $\mu_H$ = 1/($ne\rho_{xx}$), where $\rho_{xx}$ is the resistivity data from Fig. \ref{Figure2}. For all three compounds, the carrier concentrations have step-like drops with decreasing temperature, very close to where the anomalies are observed in $\rho_{xx}(T)$, $C(T)$ and $\chi(T)$, and this reduction may be due to the partial opening of a gap. The Hall mobility $\mu_H$ increases as the temperature decreases, with an observable anomaly near $T_s$. From TiPd$_2$Sn to ZrPd$_2$Sn and HfPd$_2$Sn, the overall value of $n$ increases, while $\mu_H$ decreases.

Powder XRD measurements were performed at various temperatures below 300~K to investigate the evolution of the structure below $T_s$. Upon cooling below $T_s$, clear splitting of the major diffraction peaks can be resolved, as shown in Fig. \ref{Figure4}. These indicate that $X$Pd$_2$Sn ($X$ = Ti, Zr, Hf) undergo a structural transition across $T_s$, which is consistent with the first-order characteristics shown in the resistivity and specific heat.
For all three materials, the patterns below $T_s$ can be well indexed by a tetragonal structure (space group $I4/mmm$). Rietveld refinements of the XRD data measured at the lowest temperatures were performed based on the tetragonal structure with space group $I4/mmm$ shown in Fig. \ref{Figure4}(a) \cite{SUITS1976, YIN2015JAA}, where the $X$, Pd, and Sn are situated at the $2a$, $4d$, and $2b$ Wyckoff positions, respectively.
The calculated patterns are shown in Figs. \ref{Figure4}(e)-(g), where there is good agreement of the data with the tetragonal model for all three compounds, and the fitted lattice parameters are displayed in Table. \ref{table:table1}. It is noted that structural transitions have been reported in a few RE$T_2X$ compounds with the Heusler structure, such as (Yb, Lu)Pt$_2$In \cite{GrunerIOP2014, gruner2017charge} and REAu$_2$In (RE = La-Nd) \cite{BESNUS1986ReAUIn}.

To further investigate the nature of structural phase transition, we performed single-crystal XRD measurements on the tiny single crystals selected from the surface of the arc-melted samples. We successfully obtained clear single crystal XRD data for ZrPd$_2$Sn (shown in Fig. \ref{Figure45} (a)-(c)), where the smaller FWHM in the powder XRD patterns suggests a higher degree of crystallinity compared to the other two compounds. In Fig. \ref{Figure45}(d), the simulated patterns are shown for the cubic MnCu$_2$Al-type structure (space group $Fm\bar{3}m$), which are consistent with the experimental results at 300~K displayed in Fig. \ref{Figure45}(a). At 90~K, the main reflections are well indexed by a tetragonal structure with space group $I4/mmm$, confirming the change from cubic to tetragonal symmetry upon cooling through $T_s$. Furthermore, at 90~K, additional peaks are observed experimentally corresponding to non-integer ($hkl$), as labelled in Figs. \ref{Figure45}(b) and (c), indicating the formation of a superlattice structure.
From the measurements in the ($hk0$) plane (Fig. \ref{Figure45}(b)), peaks such as (-2, 3/4, 0) and (-2, 5/4, 0) are observed, which correspond to a commensurate modulation vector of (3/4, 0, 0). While in Fig. \ref{Figure45}(c), there are further additional peaks such as (0, 2/3, 11/4), pointing to there being multiple commensurate structural modulations below $T_s$. The appearance of a superlattice structure with multiple commensurate structural modulations below $T_s$ in ZrPd$_2$Sn suggests the onset of CDW order below the structural transition.

\begin{figure*}
  	\includegraphics[angle=0,width=0.98\textwidth]{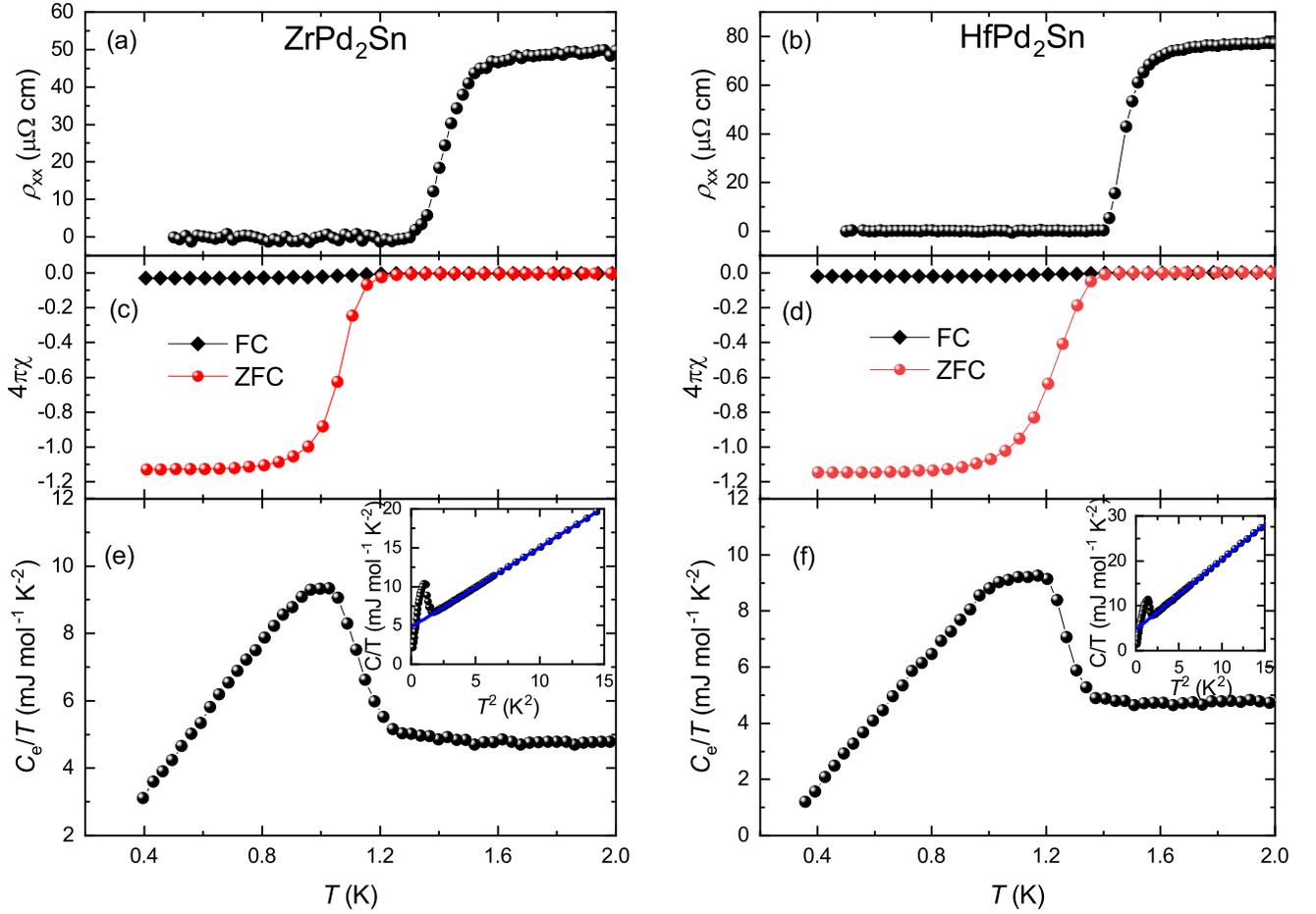}
  	\vspace{-12pt} \caption{\label{Figure5}(Color online) The temperature dependence of (a), (b): electrical resistivity and (c), (d): magnetic susceptibility and (e), (f): electronic specific heat of ZrPd$_2$Sn and HfPd$_2$Sn near the superconducting transition $T_{c}$. Insets of (e) and (f) show the total specific heat $C(T)/T$ vs $T^2$, where solid lines denote fits of the phonon contribution.  }
  \end{figure*}

    \begin{figure*}
  	\includegraphics[angle=0,width=0.98\textwidth]{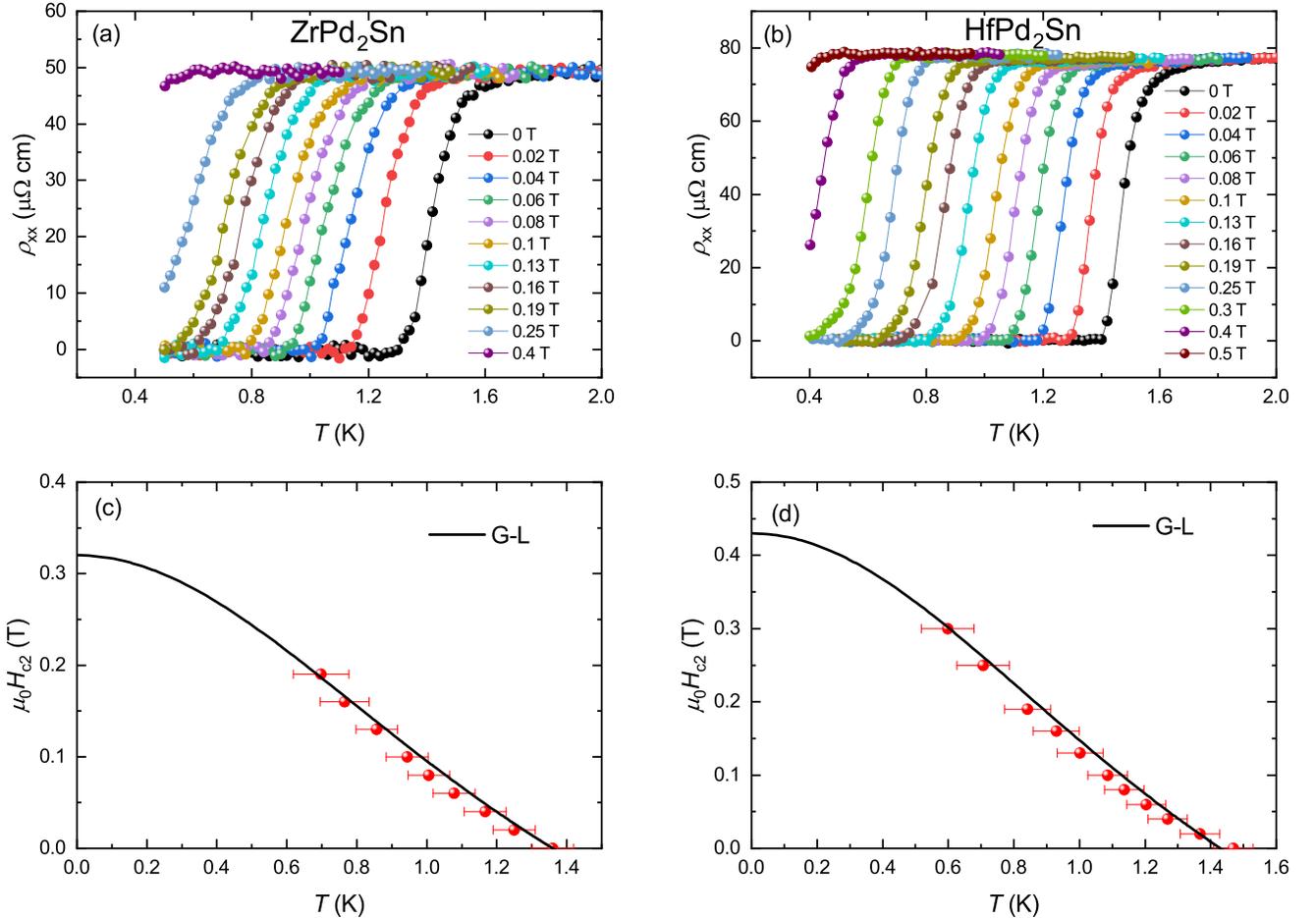}
  	\vspace{-12pt} \caption{\label{Figure6}(Color online) (a), (b): Temperature dependence of the electrical resistivity of ZrPd$_2$Sn and HfPd$_2$Sn, measured under various applied magnetic fields in the vicinity of $T_{c}$. The upper critical fields $\mu_0H_{c2}(T)$ are shown in (c) and (d), $T_c$ was determined from the midpoint of the resistivity drop at the transition. The solid lines correspond to fitting using the Ginzburg-Landau (G-L) model. }
  \end{figure*}

    \begin{figure*}
  	\includegraphics[angle=0,width=0.98\textwidth]{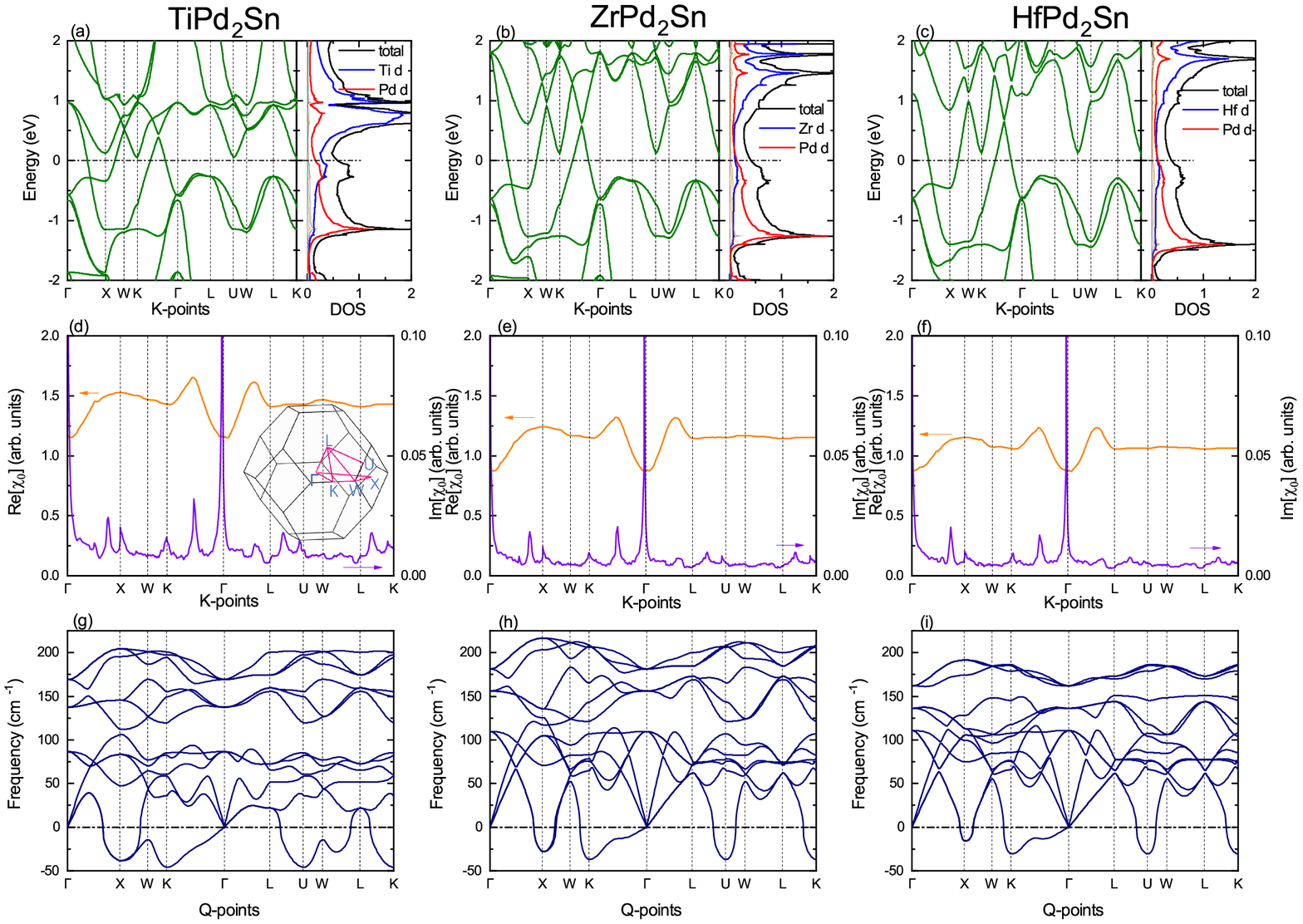}
  	\vspace{-12pt} \caption{\label{Figure7}(Color online) DFT calculations of $X$Pd$_2$Sn ($X$ = Ti, Zr, Pd). (a), (b) and (c): electronic band structure and density of states, taking into account spin-orbit coupling (SOC). (d), (e) and (f): bare electron susceptibility $\chi_0$, where the orange (purple) solid lines denote real (imaginary) parts. Inset of (d): Brillouin zone and high symmetry points, and (g), (h) and (i): Phonon spectra of the three compounds. }
  \end{figure*}
\subsection{\uppercase\expandafter{b}. Superconductivity }

Figure \ref{Figure5} displays the low temperature $\rho_{xx}(T)$, $\chi(T)$ and electronic specific heat $C_e(T)/T$ of ZrPd$_2$Sn and HfPd$_2$Sn. The resistivity $\rho_{xx}(T)$ shows sharp superconducting transitions where $\rho_{xx}$ reaches zero at around 1.2~K and 1.4~K for ZrPd$_2$Sn and HfPd$_2$Sn, respectively. The temperature dependence of $\chi(T)$ was measured under an applied magnetic field of 1~mT after zero-field cooling (ZFC) and field cooling (FC) processes. Clear diamagnetic signals are observed in the ZFC curves close to where $\rho_{xx}(T)$ reaches zero and the saturated values of $4\pi\chi$ are close to -1, which correspond to full diamagnetic shielding. The significant difference between ZFC and FC curves can be due to the magnetic flux pinning between grains. The electronic specific heat $C_e(T)/T$ are shown in Figs. \ref{Figure5}(e) and (f), after subtracting the phonon contribution $\beta T^3$, where $C_p(T)$ = $\gamma_nT+\beta T^3$ in the normal state, as shown in the insets. The fitted parameters are $\gamma_n$ = 4.77(1) and 4.72(4)~mJ~mol$^{-1}$~K$^{-2}$, $\beta$ = 1.02(1) and 1.55(1)~ mJ~mol$^{-1}$~K$^{-4}$ for ZrPd$_2$Sn and HfPd$_2$Sn, respectively. The low $\gamma_n$ values suggest weak electronic correlations. The Debye temperatures $\theta_D$ are estimated to be 197~K and 171~K, respectively, using $\theta_D=(12nR\pi^4/5\beta)^{1/3}$ with $n$ = 4 (number of atoms per formula unit) and $R$ = 8.314~J~mol$^{-1}$~K$^{-1}$ (molar gas constant). The jump of $C_e/T$ at $T_c$ indicates bulk superconductivity, where the normalized jump values $\Delta C/\gamma_nT_c$ are 1.1 and 1.2 for the respective materials. The electron-phonon coupling constant $\lambda_{\mathrm{ep}}$ is related to $T_c$ and $\theta_D$ and can be estimated using the McMillan formula
\begin{equation}\label{equation1}
\lambda_{\mathrm{ep}}=\frac{1.04+\mu^*\mathrm{ln}(\theta_D/1.45T_c)}{(1-0.62\mu^*)\mathrm{ln}(\theta_D/1.45T_c)-1.04}
\end{equation}
where using a repulsive screened Coulomb parameter $\mu^*$ of 0.13 yields $\lambda_{ep}$ values of 0.49 (ZrPd$_2$Sn) and 0.52 (HfPd$_2$Sn), indicating weak electron-phonon coupling.

Figures \ref{Figure6}(a) and (b) display the temperature dependence of the electrical resistivity of ZrPd$_2$Sn and HfPd$_2$Sn under various applied magnetic fields. The superconducting transitions are shifted to lower temperatures with increasing magnetic field. The temperature dependence of the upper critical field are shown in Figs. \ref{Figure6}(a) and (b), where the values of $T_c$ are determined from the midpoint of the resistivity drop at the transition. The data are fitted by the Ginzburg-Laudau (G-L) phenomenological model
\begin{equation}\label{equation2}
\mu_0H_{c2}(T)=\mu_0H_{c2}(0)\frac{1-(T/T_c)^2}{1+(T/T_c)^2}
\end{equation}
which yields $\mu_0H_{c2}(0)$ = 0.32(2)~T for ZrPd$_2$Sn and 0.43(2)~T for HfPd$_2$Sn. The Ginzburg-Landau coherence length $\xi_{\mathrm{GL}}$ are calculated to be 32(1)~nm (ZrPd$_2$Sn) and 28(1)~nm (HfPd$_2$Sn) via $\xi_{\mathrm{GL}}=\sqrt{\Phi_0/2\pi\mu_0H_{c2}(0)}$, where $\Phi_0$ is the magnetic flux quantum.

\subsection{\uppercase\expandafter{c}. DFT calculations}

The electronic band structure and density of states (DOS) of $X$Pd$_2$Sn ($X$ = Ti, Zr, Hf) are shown in Figs. \ref{Figure7}(a)-(c), which are consistent with the previous report \cite{WAKEELphysicaB2021}. The behaviour of the three materials are similar, where two bands crossing the Fermi level indicates a metallic nature. The $X$-d (Ti-3d, Zr-4d, Hf-5d) and Pd-4d orbitals dominate the DOS within $\pm$2~eV of the Fermi level while contributions to the DOS from other orbitals are negligible. The static bare electronic susceptibility $\chi_0$(q) are also calculated, as shown in Figs. \ref{Figure7}(d)-(f), and the imaginary parts of $\chi_0$ are obtained by calculating the nesting function. For all the three materials, the real parts of $\chi_0$ show humps in the middle of $K-\Gamma$ and $\Gamma-L$, where peaks in the imaginary parts can also be identified. In addition, several smaller nesting function peaks can also be identified around $X$, $K$ and $U$, indicating that charge fluctuations are anticipated. Figures \ref{Figure7}(g)-(i) show the phonon spectra of $X$Pd$_2$Sn calculated from first principles using the cubic Heusler structure unit cell. Imaginary phonon modes are found at $q$ = $X$, $K$ and $U$ points in all three cases, corresponding to structural instabilities and suggest the presence of a CDW \cite{ChenHuiPhysRevB2012}. Such dynamically unstable crystal structures are consistent with the experimental observations of structural phase transitions. The magnitude of the negative phonon frequencies slightly decreases from TiPd$_2$Sn to ZrPd$_2$Sn to HfPd$_2$Sn, which is consistent with the corresponding decrease in $T_s$. We note that imaginary phonon modes were also observed in LiPd$_2$Ge \cite{GornickaPhysRevB2020} and LuPt$_2$In \cite{KimPhysRevB2018, gruner2017charge}. However for LiPd$_2$Ge, the small imaginary phonon mode appears between $\Gamma$ and K and leads to superconductivity around 1~K, while for LuPt$_2$In, the large soft phonon mode corresponds to a very high CDW transition temperature around 490~K.

\begin{table}[tb]
\caption{Properties of $X$Pd$_2$Sn including the structural transition temperatures $T_s$ measured using different techniques, the lattice parameters in the cubic phase at 300~K and in the tetragonal phase at $T_{\mathrm{min}}$, as well as the Debye temperature $\theta_D$, the superconducting transition temperature $T_c$, and the zero-temperature upper critical field $\mu_0H_{c2}$(0).}
\label{table:table1}
\begin{ruledtabular}
 \begin{tabular}{c c c c}
{Parameters}             &{TiPd$_2$Sn}               &{ZrPd$_2$Sn}                &{HfPd$_2$Sn}      \\
\hline\\[-2ex]
{$T_s^{\rho\uparrow}$ (K)} &{166} &{106} &{90}     \\
{$T_s^{\rho\downarrow}$ (K)} &{158} &{102} &{86}    \\
{$T_s^{C}$ (K)} &{155} &{119} &{88}    \\
{$T_s^{\chi}$ (K)} &{181} &{115} &{95}    \\
{a (\AA) (300~K)}                                &{6.3351(3)}                &{6.5608(2)}                    &{6.5271(3)}  \\
{a (\AA) ($T_{\mathrm{min}}$)}                  &{4.5882(7)}                &{4.6622(6)}                    &{4.6155(6)}  \\
{c (\AA) ($T_{\mathrm{min}}$)}                  &{6.0451(6)}                &{6.4012(5)}                    &{6.3352(5)}  \\
{$\theta_D$ (K)} &{237} &{197} &{171}    \\
{$T_c$ (K)} &{-} &{1.2} &{1.3}    \\
{$\mu_0H_{c2}$(0) (T)} &{-} &{0.32(2)} &{0.43(2)}    \\
\end{tabular}
\end{ruledtabular}
\end{table}

\section{\uppercase\expandafter{\romannumeral4}. DISCUSSION and SUMMARY}

Our experimental measurements indicate that TiPd$_2$Sn, ZrPd$_2$Sn and HfPd$_2$Sn undergo a structural phase transition at $T_s$ $\sim$ 160 - 90~K, which potentially corresponds to a CDW instability. We confirm the existence of a superlattice structure in ZrPd$_2$Sn below $T_s$, corresponding to multiple commensurate structural modulations, suggesting the formation of CDW order. To further investigate the nature of the structural transition and possible CDW order, additional measurements are required on high quality single crystals such as electron microscopy, and x-ray scattering experiments. It is noted that superconductivity is not found in TiPd$_2$Sn down to 0.5~K, and this compound also has the largest value of $T_s$. Meanwhile, ZrPd$_2$Sn and HfPd$_2$Sn have comparatively lower values of $T_s$, and become superconducting below around 1~K. Furthermore, the signature of $T_s$ in the resistivity is more pronounced in TiPd$_2$Sn than in the other two compounds (Fig. \ref{Figure2}), which may reflect the more extensive opening of a gap on the Fermi surface for the former. These indicate possible competition between superconductivity and the CDW/structural phase transition, which has been intensively investigated in other CDW superconductors such as the dichalcogenides \cite{GabovichROPP2001} and intermetallics (Ca, Sr)$_3$(Rh, Ir)$_4$Sn$_{13}$ \cite{KlintbergPhysRevLett2012, GohPhysRevLett2015}. The origin of CDW order is often attributed to electron-phonon coupling or Fermi surface nesting. We note here that while there are peaks in the nesting function around the $q$-vectors where the imaginary phonon modes appear, these nesting function peaks are not the most prominent ones. Therefore, it is likely that the CDW in these compounds originate from electron-phonon coupling, similar to the case in Lu(Pt, Pd)$_2$In \cite{KimPhysRevB2018}. Considering that the cubic structure of $X$Pd$_2$Sn does not have low dimensional characteristics, a three-dimensional CDW is also expected. In addition, the possible existence of a CDW QCP reported in Lu(Pt$_\mathrm{1-x}$Pd$_\mathrm{x}$)$_2$In indicates that the Heusler materials are potential candidates for investigating CDW-related quantum criticality. However, different from the second-order CDW transition in LuPt$_2$In, the first-order structural phase transitions observed here in $X$Pd$_2$Sn points to the absence of quantum phase transitions or quantum criticality. On the other hand, chemical substitution and hydrostatic pressure can potentially be used to tune the structural phase transitions and superconductivity, which may provide insights into their interplay.

In summary, we have synthesized polycrystalline samples of $X$Pd$_2$Sn ($X$ = Ti, Zr, Hf) and studied their structural and physical properties. Structural phase transitions from the high temperature cubic MnCu$_2$Al-type structure ($Fm\bar{3}m$) to a low-temperature tetragonal structure, are observed in the three materials, at respective temperatures of around $T_s$ $\sim$ 160, 110, and 90~K. Low temperature single-crystal XRD measurements of ZrPd$_2$Sn show the presence of a superlattice structure below $T_s$, with multiple commensurate modulations, providing evidence for the formation of CDW order. Bulk superconductivity in ZrPd$_2$Sn and HfPd$_2$Sn is observed with respective transition temperatures $T_c$ = 1.2 and 1.3~K. DFT calculations show the presence of imaginary phonon modes that suggest a structural instability driven by electron-phonon coupling, together with humps in Fermi surface nesting functions. These findings suggest that $X$Pd$_2$Sn are promising systems for examining the interplay of superconductivity and CDW order.

\section{Acknowledgments}

We thank G. Cao, Y. Liu and S. Song for assisting with x-ray diffraction and $^3$He-SQUID measurements. This work was supported by the Key R\&D Program of Zhejiang Province, China (2021C01002), the National Natural Science Foundation of China (Grants No. 11874320, No. 11974306, No. 12034017, and No. 11874137), the Zhejiang Provincial Natural Science Foundation of China (R22A0410240), and the National Key R\&D Program of China (Grant No. 2017YFA0303100).

\end{document}